%% file: main.tex
\documentclass[conference,a4paper]{IEEEtran}

\usepackage{xcolor}
\usepackage{balance}
\usepackage{cite}
\usepackage{multirow}
\usepackage[pdftex]{graphicx}
\usepackage{amsmath,amssymb,amsfonts}
\usepackage{textcomp}
\usepackage[caption=false,font=footnotesize]{subfig}
\usepackage[nolist]{acronym}
\usepackage[capitalize]{cleveref}
\usepackage{nicefrac}
\usepackage{mathrsfs}
\usepackage{graphicx}  
\crefname{table}{Tab.}{Tabs.}
\Crefname{table}{Tab.}{Tabs.}
\crefformat{equation}{(#2#1#3)}

\begin{document}
\input{./acronyms.tex}

\title{Time-Varying Rician K-factor in Measured Vehicular Channels at cmWave and mmWave Bands}

\author{\IEEEauthorblockN{
Faruk Pasic\IEEEauthorrefmark{1},
Markus Hofer\IEEEauthorrefmark{2},
Thomas Zemen\IEEEauthorrefmark{2},
Andreas F. Molisch\IEEEauthorrefmark{3} and
Christoph F. Mecklenbr{\"a}uker\IEEEauthorrefmark{1}
}%

\IEEEauthorblockA{\IEEEauthorrefmark{1}
Institute of Telecommunications, TU Wien, Vienna, Austria}
\IEEEauthorblockA{\IEEEauthorrefmark{2}
AIT Austrian Institute of Technology, Vienna, Austria}
\IEEEauthorblockA{\IEEEauthorrefmark{3}
Ming Hsieh Department of Electrical and Computer Engineering, University of Southern California, Los Angeles, USA} \IEEEauthorblockA{faruk.pasic@tuwien.ac.at}
}

\maketitle

\begin{abstract}
Future vehicular communication systems will integrate \ac{MMW} technology to enhance data transmission rates.
To investigate the propagation effects and small-scale fading differences between \ac{MMW} and conventional \ac{CMW} bands, multi-band channel measurements have to be conducted.
One key parameter to characterize small-scale fading is the Rician \textit{K}-factor.
In this paper, we analyze the time-varying \textit{K}-factor of \ac{V2I} channels across multiple frequency bands, measured in an urban street environment. 
Specifically, we investigate three frequency bands with center frequencies of 3.2\,GHz, 34.3\,GHz and 62.35\,GHz using measurement data with 155.5\,MHz bandwidth and a sounding repetition rate of 31.25\,\textmu s.
Furthermore, we analyze the relationship between \textit{K}-factor and \ac{RMS} delay spread.
We show that the Ricean \textit{K}-factor is similar at different frequency bands and that is correlated with the \ac{RMS} delay spread.
\end{abstract}
\vskip0.5\baselineskip

\acresetall
\section{Introduction}
Due to the high utilization of conventional \ac{CMW} (sub-6\,GHz) bands, there is a bandwidth shortage that results in lower data transmission rates.
Conversely, \ac{MMW} bands (24\,GHz -- 300\,GHz)~\cite{3gpp.38.101-1} offer larger bandwidth communication channels, which enables achieving higher data rates~\cite{Ai2020}.
Therefore, \ac{MMW} technology is attracting significant attention for future vehicular communication systems. 

To understand how propagation and small-scale effects differ between \ac{CMW} and \ac{MMW} bands, comparative measurements across these frequency bands are necessary~\cite{Pasic2023_mag}.
Numerous studies in the literature have analyzed multi-band propagation in vehicular~\cite{Dupleich2019_multi, Boban2019, Hofer2021, Hofer2022, Hofer2024, Wang2020}, indoor~\cite{Ling2023, Pasic2022, Pasic2023, Radovic2023} and cellular~\cite{Miao2023} scenarios. 
These studies examine various channel parameters, such as path loss, blockage loss, angular spread, \ac{RMS} delay and Doppler spread.

One key parameter to characterize small-scale fading is the Rician $K$-factor, which defines the ratio of deterministic to stochastic multi-path components~\cite{molisch2012wireless}.
Understanding the $K$-factor is crucial for designing transmission and reception techniques aimed at mitigating the small-scale fading effects~\cite{Pasic2024, Bernado2015}.
Given the significant differences in wavelength between \ac{CMW} and \ac{MMW} bands, it is essential to examine how the $K$-factor varies across these frequency bands.
Only one measurement-based multi-band analysis of the $K$-factor can be found in the literature, specifically in~\cite{Miao2023}.
In~\cite{Miao2023}, the authors examine the $K$-factor in urban micro and outdoor-to-indoor scenarios, with configurations where either a single antenna or both the transmit and receive antennas are positioned at elevated heights within the building.
To the best of our knowledge, no other research has analyzed the $K$-factor across multiple bands.
Moreover, there is no analysis of the $K$-factor across multiple bands in vehicular scenarios, where the transmit and receive antennas are placed at average car heights.



\textbf{Contribution:}
In this paper, we present a comparative analysis of the time-varying Rician $K$-factor between the \ac{CMW} and \ac{MMW} frequency bands based on multi-band \ac{V2I} channel measurements.
Specifically, our measurements have been conducted simultaneously at center frequencies of 3.2\,GHz, 34.3\,GHz and 62.35\,GHz in an urban street environment.
Furthermore, we investigate the correlation between the $K$-factor and the \ac{RMS} delay spread.

\textbf{Organization:}
The remainder of the paper is organized as follows. 
\cref{sec:measurement} provides a detailed description of the multi-band channel measurements.
In~\cref{sec:measurement_evaluation}, we explain the evaluation and post-processing of the measured data.
\cref{sec:results} discusses the results, focusing on the Rician $K$-factor and \ac{RMS} delay spread.
Finally,~\cref{sec:conclusion} concludes the paper.

\section{Measurement Data Description} \label{sec:measurement}
\begin{figure*}
    \centering
    \includegraphics[width=\linewidth]{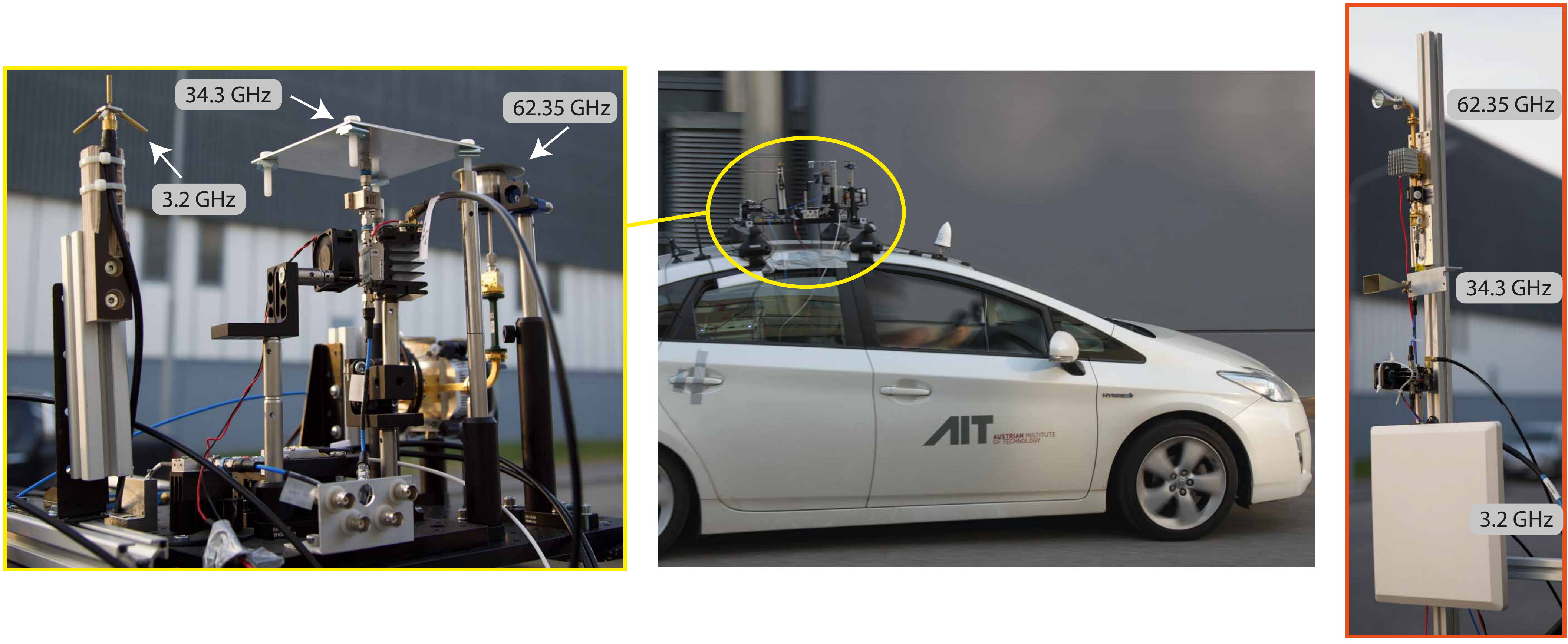}
    \caption{Transmit antennas (left) are mounted on the roof of a car (middle). Receive antennas (right) are mounted on a pole of a tripod.}
    \label{fig:setup}
\end{figure*}

For the analysis of the $K$-factor in this paper, we use \ac{V2I} multi-band channel measurements described in detail in~\cite{Hofer2021}. 
The channel measurements are conducted simultaneously at carrier frequencies of 3.2\,GHz, 34.3\,GHz and 62.35\,GHz, with a measurement bandwidth of 155.5\,MHz for each frequency band.
At the transmit side, we use custom-built omnidirectional monopole antennas mounted on a car rooftop for all three frequency bands, as shown in~\cref{fig:setup}.
The transmit antennas are arranged in a straight line along the direction of travel, with the 62.35\,GHz antenna positioned at the front, followed by the 34.3\,GHz and the 3.2\,GHz antenna.
The antennas are mounted at a uniform height to ensure that all ground planes are at the same level.
At the receive side, we use directional antennas with similar radiation patterns mounted on a tripod, as shown in~\cref{fig:setup}.
Specifically, we use a patch array antenna for 3.2\,GHz with a 17$^\circ$ \ac{HPBW} and 18\,dBi gain, a Fairview SMH128KR-20 horn antenna for 34.3\,GHz with an 18.3$^\circ$ \ac{HPBW} and 20\,dBi gain and a Pasternack PE-9881-20 conical horn antenna for 62.35\,GHz with an 18$^\circ$ \ac{HPBW} and 20\,dBi gain.
The 3.2\,GHz patch antenna is mounted 159\,cm above the ground level. 
The 34.2\,GHz and 62.35\,GHz horn antennas are mounted 34.8\,cm and 64\,cm above the 3.2\,GHz patch antenna, respectively.

\begin{figure}
    \centering
    \includegraphics[width=0.95\linewidth]{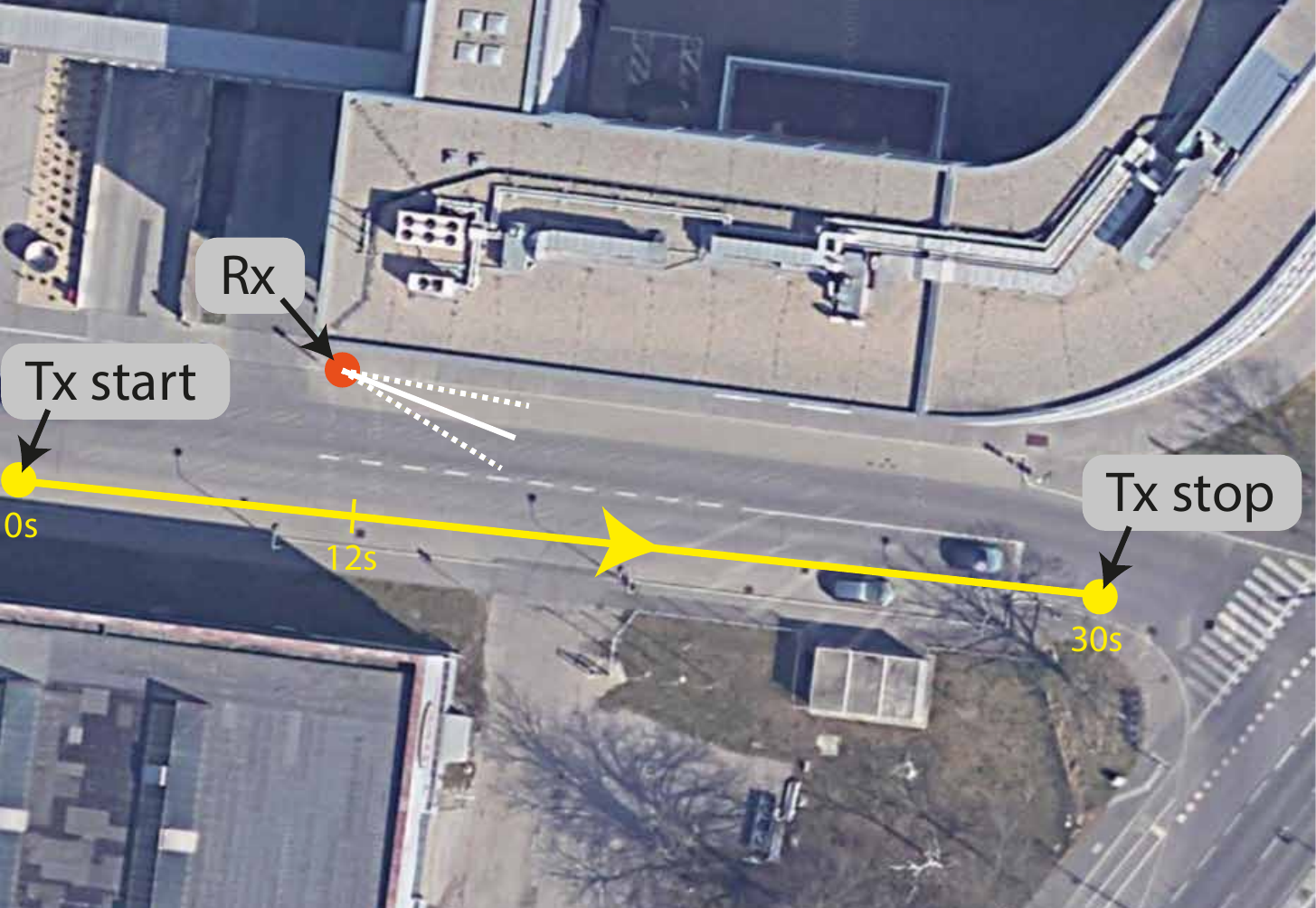}
    \caption{ A car approaches an intersection and stops. The directional receive antennas are pointed towards the intersection.}
    \label{fig:campaign}
\end{figure}

\begin{table}[t]			
	\caption{Channel Sounding Parameters}
	\begin{center}
		\label{tab:chSoundingParameters}
		\begin{tabular}{r l}
			\hline		
		    \textbf{Parameter}  & \textbf{Value} \\
		    \hline	
                Carrier Frequency $f_\mathrm{c}$                 & 3.2, 34.3, 62.35\,GHz \\
                Number of Subcarriers $K$                        & 311 \\
			Subcarrier Spacing $\bigtriangleup f$            & 500\,kHz \\
			Bandwidth $B$                                    & 155.5\,MHz \\
			Symbol Duration $t_\mathrm{s}$                   & 2\,\textmu s \\
			Symbols per Snapshot $N_\mathrm{sym}$            & 5 \\
                Interval Between Snapshots $t_\mathrm{R}$        & 31.25\,\textmu s \\
			Snapshot Duration  $t_\mathrm{snap}$             & 10\,\textmu s \\
			Number of Snapshots $N$                          & 960\,000 \\	
			Measurement Duration $t_\mathrm{m}$              & 30\,s \\
			Max. Relative Velocity $v_{\rm max}$             & 1500, 140, 77\,m/s \\
			\hline		
		\end{tabular}
	\end{center}
\end{table}

Using this setup, we conduct channel measurements in an industrial area in Vienna, as illustrated in~\cref{fig:campaign}. 
The vehicle (Toyota Prius), equipped with the transmit antennas, moves towards the receiver. 
The receiver is mounted on a tripod on the left side of the street and remains static throughout the measurement campaign.
After passing the receiver, the transmit vehicle approaches a ``T"-intersection with traffic lights and stops.
The receive antennas are oriented towards this stop point at the intersection. 

For all frequency bands, we follow the same measurement procedure.
We transmit a sequence of complex baseband \ac{OFDM} symbols with a low Crest factor~\cite{Friese1997} as the channel-sounding signal.
The measurement parameters are provided in~\cref{tab:chSoundingParameters}.
At the receiver, the measurement sequence is divided into 960\,000 snapshots, each containing 5 symbols.
The first \ac{OFDM} symbol of each snapshot is used as a cyclic prefix and then discarded. 
The remaining 4 symbols are averaged to improve the \ac{SNR}.
After \ac{OFDM} processing, we estimate the wireless channel for all subcarriers using least-square estimation given by~\cite{Loschenbrand2019}
\begin{equation}
    \mathrm{H} [k,n] = \frac{\mathrm{Y}[k,n]}{\mathrm{X}[k] \mathrm{H}^{\rm RF} [k]},
\end{equation}
where $k$ is the subcarrier index, $n$ is the snapshot index, $\mathrm{H}^{\rm RF} [k]$ represents the \ac{RF} chain calibration function, $\mathrm{X}[k]$ is the known transmit complex amplitude and $\mathrm{Y}[k,n]$ denotes the received complex amplitude.
Calibration is performed by directly connecting the transmitter with the receiver via attenuators and measuring $ \mathrm{H}^{\rm RF}[k]$.
Finally, we have the time-variant \ac{CTF} $\mathrm{H} [k, n]$, with $k \in \{ 0, \ldots, K - 1 \}$ subcarriers and $n \in \{ 0, \ldots, N - 1 \}$ time symbols.
For simplicity of notation, we omit the index for the frequency band.

\section{Measurement Evaluation} \label{sec:measurement_evaluation}
For each frequency band, we assume the channel to be locally stationary within a window of $T_{\rm stat} = $ 100\,ms of motion and over the entire frequency range, without further justification.
The chosen stationarity time-window corresponds to $N_{\rm stat} = $ 3200 time symbols per stationarity region, resulting in $L_{\rm stat} = N / N_{\rm stat} =$ 300 stationarity regions.
Hence, we have the time-variant \ac{CTF} $\mathrm{H}^{\left( i \right)} [k, n]$, with $k \in \{ 0, \ldots, K - 1 \}$ subcarriers, $n \in \{ 0, \ldots, N_{\rm stat} - 1\}$ time symbols per stationarity region and $i \in \{ 0, \ldots, L_{\rm stat} - 1 \}$ stationarity regions.
We first convert the time-variant \ac{CTF} $\mathrm{H}^{\left( i \right)} [k, n]$ to the delay domain $\mathrm{H}^{\left( i \right)} [\tau, n]$, using the \ac{IDFT}. 
The \ac{DR} is defined as the difference between the maximum power of $\mathrm{H}^{\left( i \right)} [\tau, n]$ and a level 6\,dB above the noise floor.
The noise floor is determined by the median, following the procedure described in~\cite{Sousa1994}.
To ensure a fair comparison, we choose the \ac{DR} to be the smallest one of the three bands. 
We set values smaller than the \ac{DR} to be zero.
Finally, we transform the time-variant \ac{CIR} $\mathrm{H}^{\left( i \right)} [\tau, n]$ back to the frequency domain $\mathrm{H}^{\left( i \right)} [k, n]$,  using the \ac{DFT}.

\subsection{$K$-factor Estimation}
To estimate the $K$-factor, we use the technique introduced in~\cite{Oestges2010}, based on the \ac{MoM}~\cite{Greenstein1999}.
First, we calculate the power of the time-variant channel $\mathrm{H}^{\left( i \right)} [k, n]$ as 
\begin{equation}
    P_{\mathrm{H}}^{\left( i \right)} [k, n] = 
    \bigl\lvert \mathrm{H}^{\left( i \right)} [k, n] \bigr\rvert^2.
\end{equation}
The first moment, or the average power of the time-variant channel, is given as
\begin{equation}
    \overline{P}_{\mathrm{H}}^{\left( i \right)} =
    \frac{1}{K N_{\rm stat}} 
    \sum_{k=0}^{K-1} \sum_{n=0}^{N_{\rm stat} - 1}
    P_{\mathrm{H}}^{\left( i \right)} [k, n]
\end{equation}
and the second moment of interest is the \ac{RMS} fluctuation of $P_{\mathrm{H}}^{\left( i \right)} [k, n]$ about $\overline{P}_{\mathrm{H}}^{\left( i \right)}$ given by
\begin{equation}
    {\upsilon}_{\mathrm{P}}^{\left( i \right)} =
    \sqrt{
    \frac{1}{K N_{\rm stat}} 
    \sum_{k=0}^{K-1} \sum_{n=0}^{N_{\rm stat} - 1}
    \left( P_{\mathrm{H}}^{\left( i \right)} [k, n] - \overline{P}_{\mathrm{H}}^{\left( i \right)} \right)^2
    } .
\end{equation}
Next, the power of the constant channel term is computed as
\begin{equation}
    \bigl\lvert V^{\left( i \right)}  \bigr\rvert^2 = 
    \sqrt{
    \left( \overline{P}_{\mathrm{H}}^{\left( i \right)} \right)^2 - 
    \left( {\upsilon}_{\mathrm{P}}^{\left( i \right)} \right)^2
    }
\end{equation}
and the power of the fluctuating channel term is given by
\begin{equation}
    \left( \sigma^{\left( i \right)} \right)^2 = 
    \overline{P}_{\mathrm{H}}^{\left( i \right)} - 
    \bigl\lvert V^{\left( i \right)}  \bigr\rvert^2 .
\end{equation}
Finally, the estimated $K$-factor for the $i$-th stationarity region is given by the ratio of the constant to the fluctuating channel term, expressed as
\begin{equation}
    K^{\left( i \right)} = 
    \frac{\bigl\lvert V^{\left( i \right)}  \bigr\rvert^2}
    { \left( \sigma^{\left( i \right)} \right)^2}.
\end{equation}

\subsection{RMS Delay Spread Estimation}
To estimate the \ac{RMS} delay spread, we use the approach from~\cite{Bernado2014_Delay}.
Using the time-variant \ac{CTF} $\mathrm{H}^{\left( i \right)} [k, n]$, we first estimate the \ac{LSF} given as
\begin{equation}
    \mathcal{C}^{(i)} \left[ \tau, \nu \right] = 
    \frac{1}{{IJ}} 
    \sum\limits_{w = 0}^{IJ - 1} 
    {{{\left| {{{\mathcal{H}}^{(i)}_w } 
    \left[ \tau, \nu \right]} \right|}^2}}
\end{equation}
with the Doppler index $\nu \in \left\{ -N_{\rm stat} / 2, \dots, N_{\rm stat} / 2 - 1 \right\}$ and the delay index $\tau \in \{ 0, \dots, K-1 \}$. 
The windowed frequency response is
\begin{equation}
    \begin{split}
    \mathcal{H}_w^{(i)} \left[ \tau, \nu \right] & =  
    \sum\limits_{k = - K/2}^{K/2} 
    \sum\limits_{n = - N_{\rm stat}/2}^{N_{\rm stat}/2 - 1}    
    \mathrm{H} \left[ k, n + i N_{\rm stat} \right] \\ 
    & \cdot \, \mathrm{G}_w [k, n] \, \text{e}^{ - {\text{j}}2\pi (\nu n - \tau k)},
    \end{split}
\end{equation}
where the tapers $\mathrm{G}_w [k,n]$ are two-dimensional discrete prolate spheroidal sequences as shown in detail in~\cite{Bernado2014_Delay, Slepian1978}.
The number of tapers in the time and frequency domain is set to $I = 2$ and $J = 1$~\cite{Hofer2021}, respectively.
We calculate the \ac{PDP} as the expectation of the \ac{LSF} over the Doppler domain, given by
\begin{equation}    
    \mathcal{P}^{(i)}\left[ \tau \right] = 
    \frac{1}{N_{\rm stat}} 
    \sum\limits_{\nu = - N_{\rm stat}/2}^{N_{\rm stat}/2 - 1} 
     \mathcal{C}^{(i)} \left[ \tau, \nu \right].
\end{equation} 
Further, we calculate the \ac{RMS} delay spread $\sigma_{\tau}^{(i)}$ as second-order moment of $\mathcal{P}^{(i)}\left[ \tau \right]$ given by
\begin{equation}
	\sigma_{\tau}^{(i)} = 
	\sqrt{{\sum\limits_{\forall \tau} \tau^{2} \mathcal{P}^{(i)}\left[ \tau \right]  \over \sum\limits_{\forall \tau} \mathcal{P}^{(i)}\left[ \tau \right] }  
	- \left( {\sum\limits_{\forall \tau } \tau \mathcal{P}^{(i)}\left[ \tau \right]  \over \sum\limits_{\forall \tau } \mathcal{P}^{(i)}\left[ \tau \right] } \right)^2  }.
\end{equation}
Finally, we compute the correlation coefficient between the $K$-factor $K^{\left( i \right)}$ and the \ac{RMS} delay spread $\sigma_{\tau}^{(i)}$ given by
\begin{equation}
    \rho = 
    \frac{ \sum_{i=0}^{L_{\rm stat}-1} \left( K^{\left( i \right)} - \overline{ K } \right)  
    \left( \sigma_{\tau}^{(i)} - \overline{ \sigma_{\tau} } \right) }
    {\sqrt{\sum_{i=0}^{L_{\rm stat}-1} \left( K^{\left( i \right)} - \overline{ K } \right)^2} 
    \sqrt{\sum_{i=0}^{L_{\rm stat}-1} \left( \sigma_{\tau}^{(i)} -  \overline{ \sigma_{\tau} } \right)^2}},
\end{equation}
where $\overline{ K } = \frac{1}{L_{\rm stat}} \sum_{i=0}^{L_{\rm stat}-1} K^{\left( i \right)}$ and $\overline{ \sigma_{\tau} } = \frac{1}{L_{\rm stat}} \sum_{i=0}^{L_{\rm stat}-1} \sigma_{\tau}^{(i)}$ represent the mean values of the $K$-factor and the \ac{RMS} delay spread, respectively.

\section{Results} \label{sec:results}
\begin{figure}[t]
    \centering
    \includegraphics[width=0.98\columnwidth]{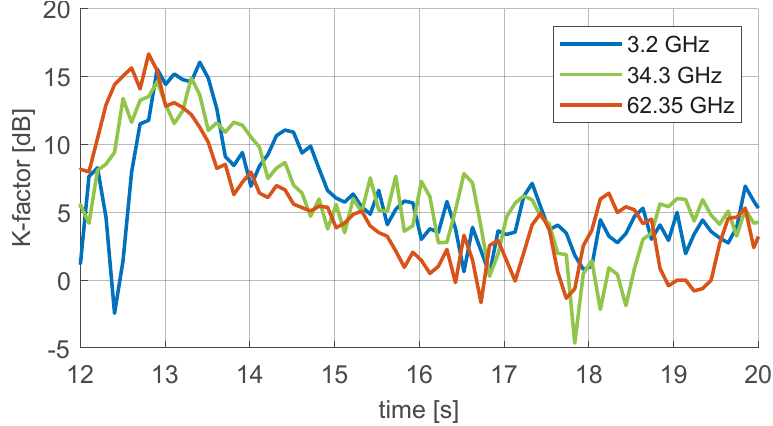} 
    \caption{The $K$-factor varies over time and exhibits a similar temporal trend for both \ac{CMW} and \ac{MMW} bands. }
    \label{fig:TV_K_factor}
 \end{figure}

\begin{figure}[t]
    \centering
    \includegraphics[width=0.98\columnwidth]{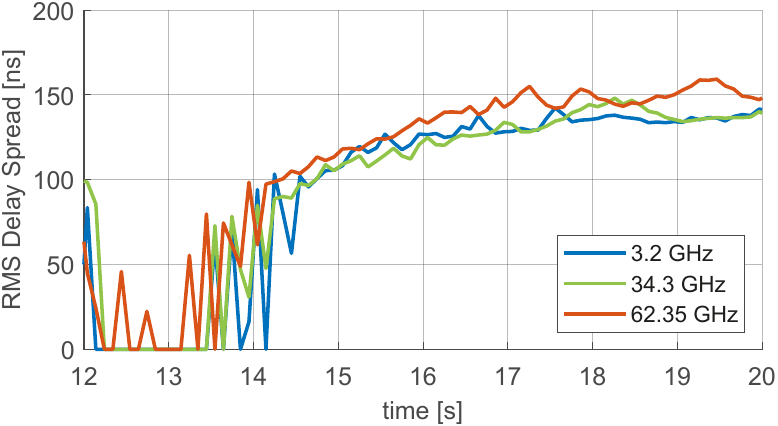} 
    \caption{The \ac{RMS} delay spread varies over time and shows a clear relationship to the $K$-factor.}
    \label{fig:TV_RMS_DS}
\end{figure}

\begin{figure}[t]
    \centering
    \includegraphics[width=0.98\columnwidth]{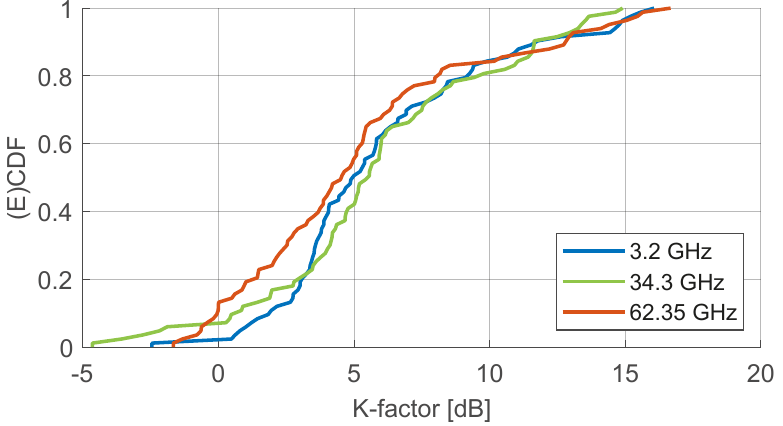} 
    \caption{In the three frequency bands, the \ac{CDF} of the $K$ factor shows similar behavior.}
    \label{fig:CDF_K_factor}
\end{figure}

\begin{figure}[t]
    \centering
    \includegraphics[width=0.98\columnwidth]{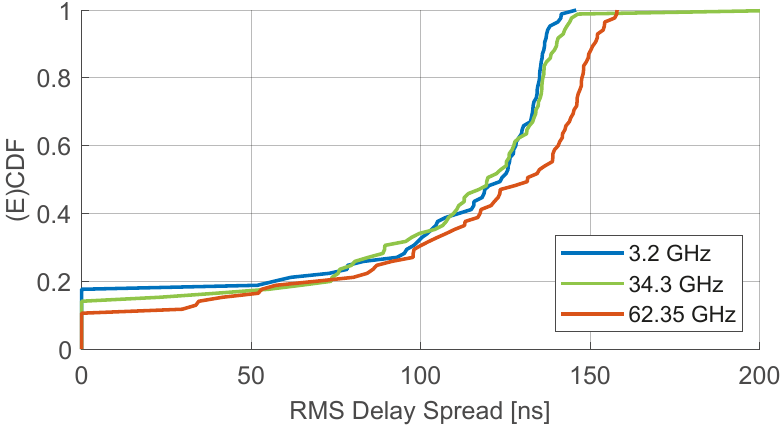} 
    \caption{The \ac{CDF} of the \ac{RMS} delay spread across \ac{CMW} and \ac{MMW} bands looks similar, with only minor variations. }
    \label{fig:CDF_RMS_DS}
\end{figure}


Here, we present the results obtained using the procedure outlined in~\cref{sec:measurement_evaluation}. 
Although the entire measurement spans 30\,s, we focus on a specific time interval.
During the first 12\,s, the transmit car is located behind the receiver, whose antennas are oriented in the opposite direction. 
As a result, the \ac{LOS} component is negligible, making the estimation of the time-varying $K$-factor irrelevant in this period.
From 12 to 20\,s, the transmit car passes the receiver and moves towards the intersection. 
At this point, the transmit car moves within the main lobe of the receive antennas. 
Finally, between 20 and 30\,s, the transmit car remains static at the intersection, leading to a loss of time variability in the channel.
Therefore, our analysis focuses on the 12 to 20\,s time span, where the most relevant channel dynamics occur.

We plot in~\cref{fig:TV_K_factor} the time evolution of the $K$-factor for three different frequency bands.
The results show that the $K$-factor varies over time and exhibits a similar temporal trend across all three frequency bands.
As the transmit car passes the receiver and enters the main lobe of the receive antennas (between 12 and 13\,s), the $K$-factor increases significantly. 
In this period, the free-space \ac{LOS} component dominates, causing the $K$-factor to reach values around 15\,dB. 
However, as the transmit car moves further away from the receiver (from 13 to 20\,s), the $K$-factor decreases. 
In this case, there are more significant diffuse components present in the received signal, leading to a reduction in the $K$-factor, which drops mainly to values between 0 and 5\,dB.

In~\cref{fig:TV_RMS_DS}, we show the time evolution of the \ac{RMS} delay spread for three different bands.
The \ac{RMS} delay spread varies over time, exhibiting a similar trend across different bands.
Between 12 and 13\,s, as the transmit car approaches the main lobe of the receive antennas, the \ac{RMS} delay spread drops to around zero. 
As the car moves away from the receiver, the \ac{RMS} delay spread increases to approximately 150\,ns. 
This demonstrates a clear relationship between the $K$-factor and the \ac{RMS} delay spread: when the $K$-factor decreases, the \ac{RMS} delay spread increases, and vice versa.


In~\cref{fig:CDF_K_factor} and~\cref{fig:CDF_RMS_DS}, we plot the \ac{CDF} of the $K$-factor and the \ac{RMS} delay spread.  
The obtained mean and standard deviation values are summarized in~\cref{tab:evaluationParameters}.
We observe that the obtained mean and standard deviation values at the three frequency bands are comparable although they do not have the same numerical value.

The estimated correlation coefficient $\rho$ between the $K$-factor and the \ac{RMS} delay spread for all three frequency bands is given in~\cref{tab:evaluationParameters}.
Additionally, we tabulated the specified correlation coefficient for the urban micro scenario from \ac{3GPP}~\cite{3gpp.38.901}.
We conclude that the estimated correlation coefficient is close to the value specified in \ac{3GPP}~\cite{3gpp.38.901}.
We observe that the correlation coefficient is negative, frequency dependent, and its magnitude increases with carrier frequency. 
We emphasize that the used antennas have approximately equal directivity and variation in correlation coefficient is not explainable by different directivity of the antennas.

\begin{table}[t]
    \caption{Estimated Parameters from Measurements and \ac{3GPP}~\cite{3gpp.38.901}}
    \begin{center}
        \begin{tabular}{c|cc|cc|c|c}
                                            & \multicolumn{2}{c|}{$K$-factor [dB]} & \multicolumn{2}{c}{ $ \sigma_{\tau} $ [ns]} & \multicolumn{2}{|c}{$\rho$} \\ \cline{2-7} 
        \multicolumn{1}{l|}{Freq. Band}     & \multicolumn{5}{c|}{Estimated}                                        & 3GPP \\ \cline{2-7} 
                                            & Mean            & Std.          & Mean       & Std.   &               &     \\ \hline
        3.2\,GHz                            & 4.93            & 3.98          & 123.66     & 49.87  &   $-$0.498    & $-$0.7  \\ \hline
        34.3\,GHz                           & 5.49            & 4.25          & 119.63     & 49.05  &   $-$0.703    & $-$0.7  \\ \hline
        62.35\,GHz                          & 4.52            & 4.48          & 131.48     & 49.46  &   $-$0.769    & $-$0.7    
        \end{tabular}
    \end{center}
    \label{tab:evaluationParameters}
\end{table}

\section{Conclusion} \label{sec:conclusion}
We analyze the time-varying Rician $K$-factor from \ac{V2I} multi-band channel measurements with a moving transmitter.
Specifically, the measurements are conducted simultaneously at center frequencies of 3.2\,GHz, 34.3\,GHz and 62.35\,GHz, using antennas with comparable radiation patterns.
The measured results show that the $K$-factor is not constant, but varies over time.
Furthermore, the results demonstrate only minor differences in the $K$-factor across different frequency bands, indicating similar propagation conditions at both \ac{CMW} and \ac{MMW} frequencies.
Moreover, the $K$-factor and \ac{RMS} delay spread are shown to be inversely related: as the $K$-factor decreases, the \ac{RMS} delay spread increases, and vice versa, consistent with existing literature.
The correlation coefficient between the $K$-factor and the \ac{RMS} delay spread is frequency dependent and its magnitude increases with carrier frequency. 

\section*{Acknowledgment}
The work of F.~Pasic was supported by the Austrian Marshall Plan Foundation with a Marshall Plan Scholarship.
The work of M.~Hofer and T.~Zemen was funded within the Principal Scientist grant Dependable Wireless 6G Communication Systems (DEDICATE 6G).
The work of A.~F.~Molisch was partly funded by the National Science Foundation.

\bibliography{references}
\bibliographystyle{IEEEtran}

\end{document}

%% file: acronyms.tex
\begin{acronym}
\acro{AWG}[AWG]{Arbitrary Waveform Generator}
\acro{CW}[CW]{Continuous Wave}
\acro{HST}[HST]{High-Speed Train}
\acro{IF}[IF]{Intermediate Frequency}
\acro{LO}[LO]{Local Oscillator}
\acro{LSF}[LSF]{local scattering function}
\acro{MMW}[mmWave]{millimeter wave}
\acro{CMW}[cmWave]{centimeter wave}
\acro{OFDM}[OFDM]{orthogonal frequency-division multiplexing}
\acro{PCB}[PCB]{Printed Circuit Board}
\acro{SMD}[SMD]{Surface Mount Device}
\acro{SNR}[SNR]{signal-to-noise ratio}
\acro{RF}[RF]{radio frequency}
\acro{V2X}[V2X]{vehicle-to-everything}
\acro{IC}[IC]{Integrated Circuit}
\acro{FPGA}[FPGA]{Field Programmable Gate Array}
\acro{ISI}[ISI]{Inter-Symbol Interference}
\acro{CSIT}{Channel State Information at the Transmitter}
\acro{ML}{maximum likelihood}
\acro{RNN}{Recurrent Neural Networks}
\acro{LSTM}{Long Short-Term Memory}
\acro{GRU}{Gated Recurrent Unit}
\acro{FDD}{frequency-division duplex}
\acro{TDD}{time-division duplex}
\acro{CIR}{channel impulse response}
\acro{CTF}{channel transfer function}
\acro{PDP}{power delay profile}
\acro{DSD}{Doppler power spectral density}
\acro{IFFT}{Inverse Fast Fourier Transform}
\acro{ITS}[ITS]{intelligent transportation systems}
\acro{5G}[5G]{fifth generation}
\acro{NR}[NR]{new radio}
\acro{QAM}[QAM]{Quadrature Amplitude Modulation}
\acro{ICI}[ICI]{Inter-Carrier Interference}
\acro{MSE}[MSE]{mean square error}
\acro{BER}[BER]{bit error ratio}
\acro{RMS}[RMS]{root-mean-square}
\acro{TDL}[TDL]{Tapped Delay Line}
\acro{ISI}[ISI]{Inter-Symbol Interference}
\acro{DFT}[DFT]{discrete Fourier transform}
\acro{IDFT}[IDFT]{inverse discrete Fourier transform}
\acro{CTF}[CTF]{channel transfer function}
\acro{DR}[DR]{dynamic range}
\acro{HPBW}[HPBW]{half-power beamwidth}
\acro{DPSS}[DPSS]{discrete prolate spheroidal sequences}
\acro{CDF}[CDF]{cumulative distribution function}
\acro{PDF}[PDF]{probability density function}
\acro{MoM}[MoM]{method of moments}
\acro{V2I}[V2I]{vehicle-to-infrastructure}
\acro{LOS}[LOS]{line-of-sight}
\acro{3GPP}[3GPP]{3rd Generation Partnership Project}
\end{acronym}